\documentclass[pra,superscriptaddress,showpacs,preprint,aps]{revtex4-1}

\usepackage{epsfig}             
\usepackage{hyperref}           
\usepackage{color}
\usepackage{colortbl}
\usepackage{graphicx}
\usepackage{amssymb,amsmath}
\begin{document}

\title{Formation of very low energy states crossing the ionization threshold of argon atoms in strong mid-infrared fields}

\author{Benjamin Wolter}
\email[]{benjamin.wolter@icfo.eu}
\affiliation{ICFO - Institut de Ci\`{e}ncies Fot\`{o}niques, Mediterranean Technology Park, 08860 Castelldefels (Barcelona), Spain}

\author{Christoph Lemell}
\email[]{lemell@concord.itp.tuwien.ac.at}
\affiliation{Institute for Theoretical Physics, Vienna University of Technology, Wiedner Hauptstr.\ 8-10/E136, A-1040 Vienna, Austria, EU}

\author{Matthias Baudisch}
\affiliation{ICFO - Institut de Ci\`{e}ncies Fot\`{o}niques, Mediterranean Technology Park, 08860 Castelldefels (Barcelona), Spain}

\author{Michael G. Pullen}
\affiliation{ICFO - Institut de Ci\`{e}ncies Fot\`{o}niques, Mediterranean Technology Park, 08860 Castelldefels (Barcelona), Spain}

\author{Xiao-Min Tong}
\affiliation{Center for Computational Sciences, University of Tsukuba, Ibaraki 305-8577, Japan}

\author{Micha\"{e}l Hemmer}
\affiliation{ICFO - Institut de Ci\`{e}ncies Fot\`{o}niques, Mediterranean Technology Park, 08860 Castelldefels (Barcelona), Spain}

\author{Arne Senftleben}
\affiliation{Institute of Physics, Center for Interdisciplinary Nanostructure Science and Technology (CINSaT), University of Kassel, Heinrich-Plett-Strasse 40, 34132 Kassel, Germany}

\author{Claus Dieter Schr\"oter}
\affiliation{Max-Planck-Institut f\"ur Kernphysik, Saupfercheckweg 1, 69117 Heidelberg, Germany}

\author{Joachim Ullrich}
\affiliation{Max-Planck-Institut f\"ur Kernphysik, Saupfercheckweg 1, 69117 Heidelberg, Germany}
\affiliation{Physikalisch-Technische Bundesanstalt, Bundesallee 100, 38116 Braunschweig, Germany}

\author{Robert Moshammer}
\affiliation{Max-Planck-Institut f\"ur Kernphysik, Saupfercheckweg 1, 69117 Heidelberg, Germany}

\author{Jens Biegert}
\affiliation{ICFO - Institut de Ci\`{e}ncies Fot\`{o}niques, Mediterranean Technology Park, 08860 Castelldefels (Barcelona), Spain}
\affiliation{ICREA - Instituci\'{o} Catalana de Recerca i Estudis Avan\c{c}ats, 08010 Barcelona, Spain}
\affiliation{Kavli Institute for Theoretical Physics, University of California, Santa Barbara, CA 93106-4030, USA}

\author{Joachim Burgd\"orfer}
\affiliation{Institute for Theoretical Physics, Vienna University of Technology, Wiedner Hauptstr.\ 8-10/E136, A-1040 Vienna, Austria, EU}
\affiliation{Kavli Institute for Theoretical Physics, University of California, Santa Barbara, CA 93106-4030, USA}

\date{\today}

\begin{abstract}
Atomic ionization by intense mid-infrared (mid-IR) pulses produces low electron energy features that the strong-field approximation, which is expected to be valid in the tunneling ionization regime characterized by small Keldysh parameters ($\gamma \ll 1$), cannot describe. These features include the low-energy structure (LES), the very-low-energy structure (VLES), and the more recently found zero-energy structure (ZES). They result from the interplay between the laser electric field and the atomic Coulomb field which controls the low-energy spectrum also for small $\gamma$. In the present joint experimental and theoretical study we investigate the vectorial momentum spectrum at very low energies. Using a reaction microscope optimized for the detection of very low energy electrons, we have performed a thorough study of the three-dimensional momentum spectrum well below 1 eV. Our measurements are complemented by quantum and classical simulations, which allow for an interpretation of the LES, VLES and of the newly identified ZES in terms of two-dimensional Coulomb focusing and recapture into Rydberg states, respectively.
\end{abstract}
\maketitle

\section{Introduction}
In the strong-field regime of atomic ionization characterized by small Keldysh parameters $\gamma=\sqrt{I_p/2U_p}\ll 1$ \cite{Keldysh} (where $I_p$ is the atomic ionization potential, $U_p=F_0^2/4\omega^2$ the ponderomotive energy, $F_0$ the field amplitude, and $\omega$ the laser frequency), the laser field is expected to dominate over the Coulomb field of the ionized atom. Accordingly, in the strong-field approximation (SFA) the Coulomb field is neglected in the final continuum state of the outgoing electron. With the recent availability of intense long wavelength ($\lambda\geq 2\, \mu$m) laser pulses, this picture has fundamentally changed: the experimental energy
spectrum of ionized electrons is found to have unexpected features at a few eV \cite{Hickstein2012,Moeller2014}, with a particular example being the low-energy structure (LES) \cite{Blaga,Quan2009}. These results strongly deviate from SFA predictions even though they were observed well within the tunneling regime ($\gamma\ll 1$). They can be explained in terms of multiple rescattering of the ionized electron at the Coulomb field of the residual ion \cite{Liu,Lemell_PRAR,Chu,Kastner1,Kastner2,Lemell_LES,Guo,Lin,Tong2013} after one or more excursions out to large distances with quiver amplitude $\alpha= F_0/\omega^2\propto F_0\lambda^2$. Recently, additional contributions at even lower energies, termed very-low-energy structures (VLESs), were discovered \cite{Wu2012} but up to now these VLESs have not been quantitatively analyzed and characterized. Our previous measurement of the three-dimensional momentum distributions of various atomic and molecular targets \cite{Dura_srep,Pullen} confirmed signatures in momentum space that correspond to the VLES and additionally revealed a zero-energy structure (ZES). Up to now, the origin of the ZES has remained speculative, with depletion of Rydberg states \cite{Dura_srep,Pullen} being a suggested mechanism. On a more fundamental level, the correspondence between quantum and classical dynamics very close to the ionization threshold is of concern as most models describing the distant rescattering scenario are based on classical dynamics. In particular, the explanations of the LES in terms of one-dimensional (transverse direction only) \cite{Kastner1,Kastner2} or two-dimensional (longitudinal and transverse directions) focusing of the emitted electron near the inner turning point \cite{Lemell_PRAR,Lemell_LES} rely on subtle properties of the classical phase space. Its applicability to the VLES and even ZES is not obvious as the deBroglie wavelength diverges for electronic energies $E$ approaching the threshold, $E\to 0$.

In the present joint experimental and theoretical study we analyze the behavior of the LES, VLES, and ZES as a function of pulse duration of the mid-infrared (mid-IR) pulse varying between 41 fs (4 optical cycles) to 140 fs ($\sim 13$ optical cycles). As the number of distant recollisions with the ionic core is controlled by the number of optical cycles and the peak-to-peak amplitude after tunneling ionization, variation of the pulse duration provides a knob to manipulate the focusing properties of the recollision process. The vectorial momentum distribution parallel ($p_\|$) and perpendicular ($p_\perp$) to the laser polarization axis is recorded by a reaction microscope (ReMi) \cite{Moshammer1996,Ullrich2003} that is optimized for low electron energies. The detection of the distribution in all three spatial dimensions allows the influence of laser pulse duration on the low-energy features to be investigated. In parallel, classical trajectory Monte Carlo (CTMC) as well as time-dependent Schr\"odinger equation (TDSE) calculations permit the identification of the VLES in terms of high-order LES contributions, i.e., two-dimensional focusing after multiple returns into phase-space regions of very high angular momenta. In addition, the ZES is unambiguously identified in terms of electrons recaptured into high Rydberg states, often referred to as frustrated field ionization \cite{Yosh,Nubbe,Eich}, which are subsequently ionized by the static electric field of the ReMi after the end of the pulse. We find excellent quantitative agreement between experiment and theory. Moreover, classical simulations reproduce many features of the quantum calculation remarkably well.

Atomic units (a.u., $\hbar=m_e=e=1$) are used unless indicated otherwise.

\section{Methods}
\subsection{Experiment}
The experiments were performed using a home-built optical parametric chirped pulse amplification (OPCPA) source \cite{Hemmer2013a,Hemmer2013} which delivers linearly polarized few-cycle pulses in the mid-IR ($\lambda= 3100$ nm center wavelength) and 160 kHz pulse repetition rate. After compression, pulse energies of up to 18 $\mu$J are achieved with a power stability of better than 1\%\ rms over 4.5 hours \cite{Thai2011}. The OPCPA is optically carrier-to-envelope phase (CEP) self-stabilized and routinely achieves 250 mrad stability over 11 minutes. The measurements were performed at a fixed laser intensity of $I_0=0.9\times10^{14}$ W/cm$^{2}$ corresponding to a Keldysh parameter of $\gamma=0.31$ and a ponderomotive energy of $U_p=80.8$ eV. The experiment was conducted for pulse durations of 4 cycles [$\tau_p =41$ fs, full width at half maximum (FWHM) of the intensity], 6.5 cycles ($\tau_p =68$ fs) and 13 cycles ($\tau_p =140$ fs).

A ReMi \cite{Moshammer1996,Ullrich2003} is used to detect the fragments of the photoionization process. In a ReMi detection system, static electric and magnetic fields are combined to guide all charged fragments of the ionization process onto two opposing micro channel plate detectors. Combining the time-of-flight (TOF) and spatial informations, we can retrieve the 3D momentum vectors (with $\hat{e}_z$ as the direction of the static fields and the laser polarization, $\hat{e}_x$ as the laser propagation direction, and $\hat{e}_y$ as the direction of the gas jet) and the kinetic energy of the ions and electrons with very high resolution. In this work, the static fields of the ReMi were set to $F_\mathrm{extr} = 1.3$ V/cm and $B = 4.4$ Gs which results in an electron momentum accuracy of $\Delta p_\|=0.007$ a.u.\ parallel and $\Delta p_\perp=0.013$ a.u.\ transverse to the laser polarization direction for the evaluated data. Additionally, it is possible to detect the particles in coincidence, which allows the assignment of every detected photo-electron to its parent ion. The argon gas is supersonically expanded into ultra-high vacuum. In the experiment we purposely averaged over the carrier-envelope phase $\varphi_\mathrm{CEP}$.

\subsection{Theory}
We perform both quantum and classical simulations in order to probe the degree of quantum--classical correspondence at very low electron energies which is, \textit{a priori}, not obvious. This point is highlighted by the fact that the deBroglie wavelength of a typical low-energy electron is quite large, $\lambda_e(E =1\ \mathrm{eV})\approx 25$ a.u.\ $\approx 1.25$ nm.

Quantum simulations of the interaction of strong laser pulses with atoms in all dimensions remains a challenge in the mid-IR region. This is because the quiver amplitude, and therefore the excursion from the ionic core, is on the order of $\alpha=F_0/\omega^2\approx 250$ a.u., which requires a large domain in configuration space to be accurately calculated. Technical details of our three-dimensional (3D) quantum-mechanical simulation have been described previously \cite{Tong1,Tong2}. In brief, we discretize the radial coordinate space in a pseudo-spectral grid and propagate the wave function which is expanded into a set of pseudostates with quantum numbers $\tilde n, \ell, m$ employing the velocity gauge for the coupling to the laser field. To prevent unphysical reflections from the ``boundaries'' of the computing box we separate the real space into inner ($R<R_c$) and outer ($R_c<R<R_\mathrm{max}$) regions. The parts of the wavefunction reaching the outer region are projected onto Coulomb-Volkov states in momentum space and are propagated analytically henceforth. For the calculations presented here we chose an outer box radius of $R_\mathrm{max} = 1250$ a.u.\ and an inner radius of $R_c = 750$ a.u. $R_c$ needs to be significantly (at least a factor of two) larger than the quiver radius. Moreover, the value of $R_c$ determines the range of pseudostates which closely resemble atomic states with principal quantum number $n$, $\langle r\rangle_{\tilde n}<R_c$. In the present case, Rydberg states up to $n\approx 20$ are faithfully accounted for. Higher lying bound states are included in terms of pseudostates with $\ell < 80$ where the hydrogenic $\ell$ degeneracy is lifted. This allows for a coarse-grained representation of the spectral excitation density $\rho(E)$ close to the continuum threshold at $E=0$. $\rho(E)$ plays an important role in representing the process of recapture into Rydberg states, which is the key to the ZES. We describe the ionization of the Ar3$p$ electron in terms of a model potential that reproduces $I_p$ and the oscillator strengths of the coupling to the first excited states (4$d$, 5$d$) \cite{Tong3}.

In our CTMC simulation we employ an effective potential derived from the Hartree-Fock (HF) approximation \cite{CFF}. The binding energy $E_\mathrm{bind}$ determining the energy shell of the microcanonical ensemble was set to the HF result of $-16.0824$ eV (experimental value $E_\mathrm{bind}=-15.76$ eV \cite{NIST}). This leads to a Keldysh parameter $\gamma=\omega\sqrt{-2E_\mathrm{bind}}/F_0\approx 0.32\ll 1$ \cite{Keldysh} suggesting a strong dominance of tunneling over multiphoton ionization for the laser parameters used in the experiment. Since the electron emission originates from the Ar3$p$ subshell with quantum numbers $n=3$, $\ell = 1$, and $m=0,\pm 1$, we use the $\ell,m$-dependent ADK rates \cite{ADK}
\begin{eqnarray}
w_{m=0}&=&\left(\frac{3e}{\pi}\right)^{3/2}\frac{Z^2}{n^\ast{}^{9/2}}\left(\frac{4eZ^3}{Fn^\ast{}^4}\right)^{2n^\ast-3/2}\exp\left(-\frac{2Z^3}{3n^\ast{}^3F}\right) \label{eq1}\\
w_{|m|=1}&=&\left(\frac{3e}{\pi}\right)^{3/2}\frac{2eZ^2}{n^\ast{}^{11/2}}\left(\frac{4eZ^3}{Fn^\ast{}^4}\right)^{2n^\ast-5/2}\exp\left(-\frac{2Z^3}{3n^\ast{}^3F}\right)\,,\label{eq2}
\end{eqnarray}
where $n^\ast$ is the effective principal quantum number derived from $E_\mathrm{bind} = -Z^2/2n^\ast{}^2$. Up to constant factors, the total emission rate is therefore proportional to
\begin{equation}
w_\mathrm{total}\propto F^{3/2-2n^\ast}\left( 1+\frac{Fn^\ast{}^3}{Z^3}\right)\exp\left(-\frac{2Z^3}{3n^\ast{}^3F}\right)\, ,
\end{equation}
and the probability for emission during a time interval $\Delta t$ proportional to
\begin{equation}
P(\Delta t) = 1-\exp(-w_\mathrm{total}\Delta t)\, .
\end{equation}
Integration of $P(\Delta t)$ over the pulse duration gives the total emission probability for an atom in a laser pulse with maximum field strength $F$.

The $z$ coordinate of the tunnel exit, $z_0$, was calculated from the field strength at ionization time. The perpendicular position was taken from projections of the corresponding HF wavefunctions with $m=0$ and $|m|=1$ onto the plane perpendicular to the polarization axis at $z_0$. Widths of the Gaussian momentum distributions along the polarization direction at the tunnel exit 
\begin{equation}
\sigma^2(p_\|) = \frac{3\omega}{2\gamma^3}
\end{equation}
and the perpendicular direction
\begin{equation}
\sigma^2_\mathrm{ADK}(p_\perp) = \frac{F/2}{\sqrt{-2E_\mathrm{bind}}}\, .
\label{eq8}
\end{equation}
were derived in Ref.\ \cite{DK}. Recently, it was suggested that the transverse width is somewhat underestimated by Eqn.\ \ref{eq8} \cite{Murray,Lein,Corkum}. We have taken this correction into account by using $\sigma(p_\perp)=\sqrt{2}\sigma_\mathrm{ADK}(p_\perp)$ \cite{Murray}. For each set of initial conditions taken from properly normalized distribution functions, the trajectory is calculated in the combined fields of the remaining laser pulse and the ionic core. The propagation was performed using a standard fourth order Runge-Kutta algorithm and proceeds to electron-core distances of at least $r\geq 5000$ a.u.\ after the end of the laser pulse. Since the experimental conditions of the ReMi were set to use an extraction field of $F_\mathrm{extr}=1.3$ V/cm $\approx 2.5\times 10^{-10}$ a.u., this very weak external field was also taken into account. It is further used in the data evaluation of the CTMC simulation as it is crucial for the analysis of the ZES. For each carrier-envelope phase ($\Delta\varphi_\mathrm{CEP}=15^\circ$), $2.5\times 10^6$ trajectories were calculated and combined for the determination of the final vectorial momentum and energy vs.\ angular momentum distributions. Momentum, energy, and angular momenta were also recorded for electrons with energies $E_\mathrm{final}<0$ after the end of the pulse, signifying electrons recaptured into high-lying Rydberg states.

To accurately simulate the experiment we perform an average over the focal intensity distribution. To this end we have assumed a cylindrically symmetric beam with Gaussian intensity profile and maximum intensity $I_0$. The volume with an intensity larger than $I_i$ is given by \cite{Augst,Post}
\begin{equation}
V(I > I_i) = \frac{2\pi d_\circ^2z_R}{3} (2\beta+\beta^3/3-2\arctan\beta)
\end{equation}
with $\beta=\sqrt{I_0/I_i-1}$ and $d_\circ$ the focal diameter. This expression was evaluated for small steps of the maximum electric field ($F_{i+1}(t) = F_i(t)/1.005$) resulting in thin shells of volume
\begin{equation}
V = V(I > I_{i+1})-V(I > I_i)
\end{equation}
with maximum local field strength $F$ between $F_{i+1}$ and $F_i$.

\section{Results and Discussion}
Previously presented experimental data for the full 3D momentum spectra of electrons emitted from argon, oxygen \cite{Dura_srep} and nitrogen \cite{Pullen} interacting with ultrashort, mid-IR laser pulses allowed to identify the VLES \cite{Quan2009,Wu2012,Lemell_LES} in the momentum plane $(p_\perp = \sqrt{p_x^2 + p_y^2}$, $p_\| = p_z)$, where it appears as a V-shaped distribution. In addition, a peak at zero energy (within the energy resolution) dubbed ZES was found. The origin of these structures and the precise connection to the previously observed LES remained unclear. In the present measurement we extend the 3D momentum spectrum for the case of argon to larger $p_\|$ to simultaneously resolve several orders of low-energy structures as predicted to exist \cite{Lemell_LES}. Moreover, by varying the pulse duration at constant $\gamma$ we can control the V-shape, i.e. the opening angle, which aids in the identification of the underlying recollision and Coulomb focusing processes.

\begin{figure}[h!]
	\centering
		\includegraphics[width=0.8\textwidth]{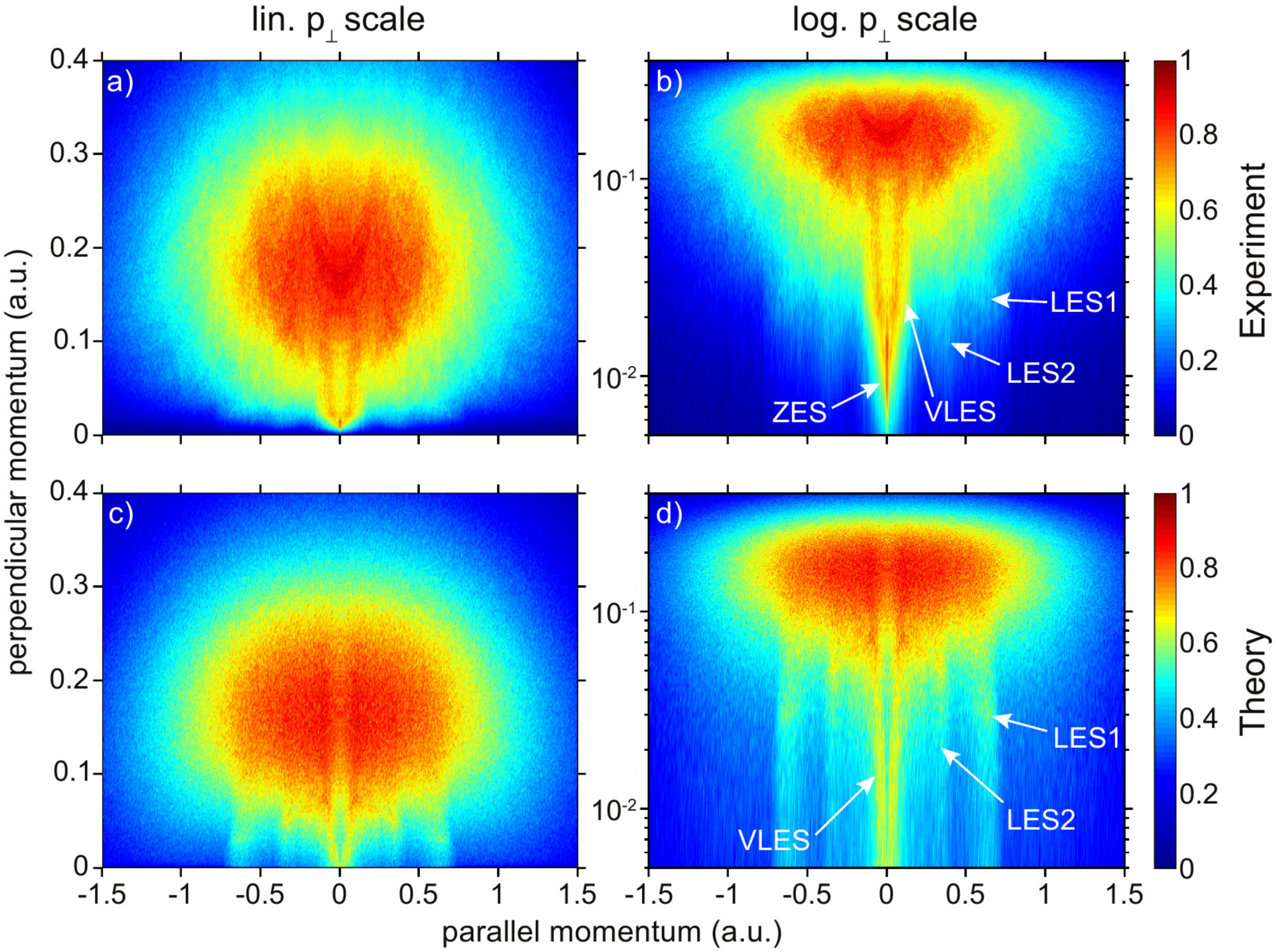}
	\caption{(Color online) Measured (top) and simulated (bottom) electron momentum distributions (normalized linear color scale) for argon interacting with a 68 fs pulse with peak intensity of $I_0=0.9\times 10^{14}$ W/cm$^2$ and a central wavelength of $\lambda =3100$ nm with $p_\perp$  on linear (left column) and logarithmic scales (right column). The first and second order LES (LES1, LES2), the VLES, and the ZES structures are marked.\label{fig:PtPl_68fs}}
\end{figure}

In Fig.\ \ref{fig:PtPl_68fs} we compare the measured and simulated electron momentum maps produced by the 6.5-cycle 68 fs mid-IR pulses [with $p_\perp$ on linear (left) and logarithmic (right) scales]. The V-shaped structure can clearly be recognized for electrons with $|p|<0.1$ a.u. in both the experimental (top) and simulated (bottom) momentum distributions. The ZES ($p_\perp\approx 0.01$ a.u.) is clearly visible in the experimental data (Fig.\ \ref{fig:PtPl_68fs}a, b). In CTMC calculations (Fig.\ \ref{fig:PtPl_68fs}c, d) considering only direct ionization to the continuum, the ZES is absent. We associate this feature with post-pulse ionization of Rydberg states, i.e., electrons with negative final energy in our CTMC simulation (see below). Additionally, lobes of the distribution towards lower transverse momenta are visible for $p_\|\approx\pm 0.4$ a.u.\ and $p_\|\approx\pm 0.6$ a.u. These lobes can be identified as LESs of first and second order by comparison with the CTMC simulation. In \cite{Lemell_LES} LESs of increasing order were attributed to turning points of the electron trajectory close to the ionic core after an increasing number of quiver oscillations in the driving laser field. Consequently, only longer laser pulses with many cycles driving the liberated electron back to the core multiple times facilitate a high-order LES.

\begin{figure}[h!]
	\centering
	  \includegraphics[width=0.5\textwidth]{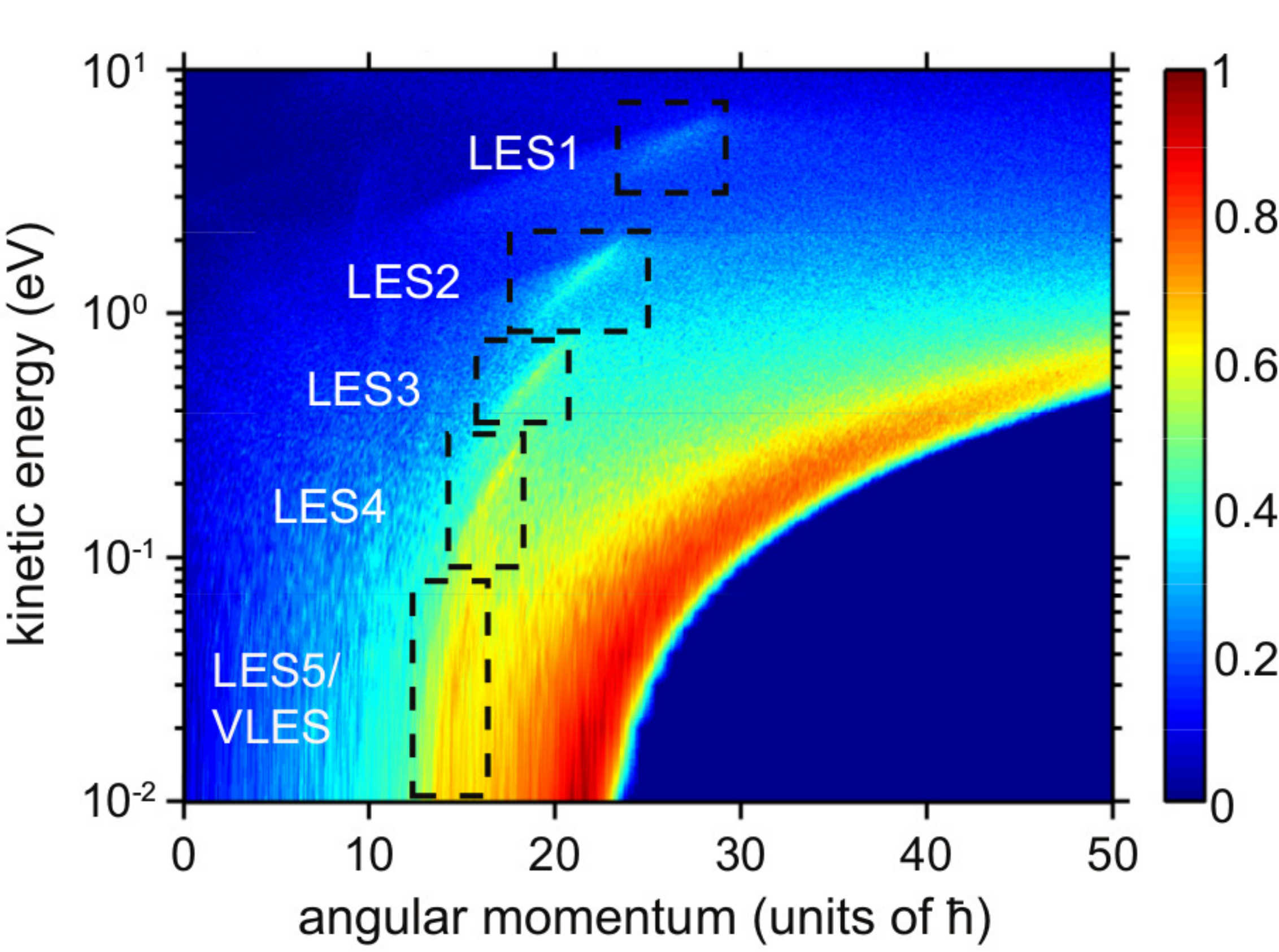}
	\caption{(Color online) Projection of the electron momentum distribution of Fig.\ \ref{fig:PtPl_68fs}c, d onto the $\log E-L$ plane (normalized linear color scale). The $k$-th order LES peaks ($k=1,\ldots,5$) resulting from two-dimensional Coulomb focusing are marked.\label{e-l}}
\end{figure}

\begin{figure}[h!]
	\centering
	  \includegraphics[width=0.8\textwidth]{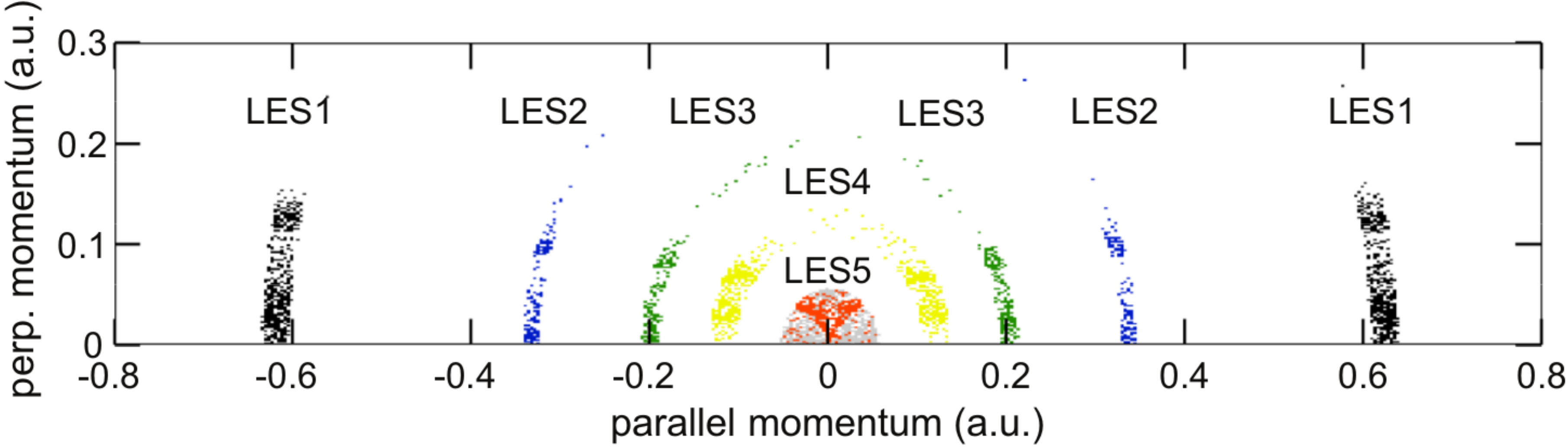}
	\caption{(Color online) Mapping of the trajectories focused within the $k^{th}$ order LES in the $E-L$ plane (dashed boxes in Fig.\ \ref{e-l}) onto the $(p_\|,p_\perp)$ plane. The near vertical stripes correspond to the low-order LES [$k=1$ (black), 2 (blue), 3 (green), 4 (yellow)] while the highest order ($k=5$) visible in Fig.\ \ref{e-l} gives rise to a V-shaped distribution (``VLES'', red). The feature-less half-circle around zero momentum corresponds to very low energy electrons outside the VLES region (gray, see text).\label{VLES_mom}}
\end{figure}

The association of low-energy structures with two-dimensional Coulomb focusing is corroborated by the projection of the final states of the trajectories in the energy--angular momentum plane. As in Ref. \cite{Lemell_LES} we can link the structures appearing at small perpendicular momenta in Fig.\ \ref{fig:PtPl_68fs} to local maxima in the energy--angular momentum distribution plotted in Fig.\ \ref{e-l}. 
Peaks due to focusing up to the fifth order of recollision can be identified (LES1 -- LES5). Their influence on the vectorial momentum spectrum can now be highlighted by selectively mapping these peaks onto the $(p_\|,p_\perp)$ plane thereby suppressing the smooth background. The resulting reduced momentum map is plotted in Fig.\ \ref{VLES_mom}.
It thus becomes obvious that the approximately vertical bands signify the low-order LES (LES1 -- LES4). The V-shaped structure at very low energy (red data points in Fig.\ \ref{VLES_mom}), which was previously identified as the VLES, corresponds to the highest order (in this case fifth order) LES. To show the difference in final momentum between VLES and other electrons with similar kinetic energies we have also plotted the final momenta of trajectories from the same energy range as the VLES but higher angular momentum ($L\sim 22$, absolute maximum of the $E-L$ distribution). This leads to the feature-less uniformly distributed background half-circle (grey data points in Fig.\ \ref{VLES_mom}) around zero momentum.

The VLES differs from lower order LESs not only by its very low energy but also by a different vectorial momentum distribution. This is corroborated by analyzing the Kepler orbits of the outgoing trajectories with eccentricities ranging from $\varepsilon=\sqrt{1+2EL^2}\approx 1$ close to threshold (VLES) to $\varepsilon\approx 17.3$ in the center of LES1. Such hyperbolas have asymptotes with inclination angles relative to the major axis ranging from $\theta=\arctan\sqrt{\varepsilon^2-1} = 0$ to almost $\theta\approx \pi/2$. As the major axes of LES electrons of all orders have similar tilt angles close to $\pi/2$ with respect to the laser polarization direction, increasing order of LES is associated with larger emission angles of the electrons reaching near perpendicular emission for VLES electrons.

\begin{figure}[htbp]
  \centering
    \includegraphics[width=0.8\textwidth]{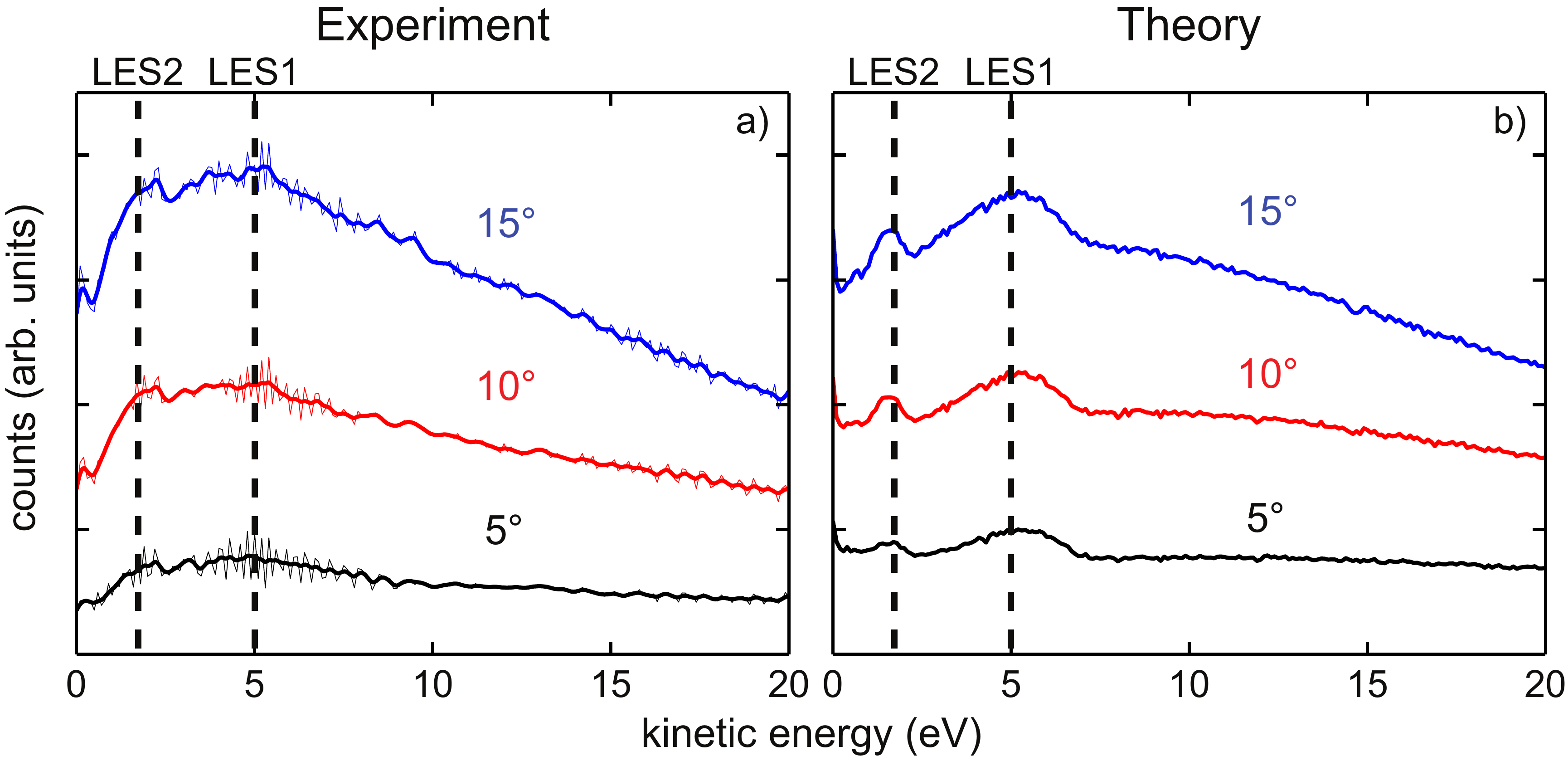}
  \caption{(Color online) Energy spectra resulting from the momentum distributions of Fig.\ \ref{fig:PtPl_68fs} for electrons emitted into a cone with opening angles $\varphi=5^\circ$, $10^\circ$, and $15^\circ$ around the polarization axis. Structures near 5 and 1.75 eV are attributed to first and second order LES, respectively (experimental raw data: thin lines, smoothed spectra: thick lines).\label{fig:KED_detectionAngle}}
\end{figure}

In order to highlight similarities and differences to the original LES observed in \cite{Blaga,Quan2009}, we zoom in on the LES and extract the kinetic energy distribution of emitted electrons for the case of 68 fs pulses by restricting the data set to detection aperture angles of $5^\circ$, $10^\circ$, and $15^\circ$ (see Fig.\ \ref{fig:KED_detectionAngle}) which are comparable to the opening aperture in a typical TOF. We attribute the structures appearing in the experiment near 5 and 1.75 eV (Fig.\ \ref{fig:KED_detectionAngle}a) to first and second order LES, respectively, which is supported by our simulated spectra (Fig.\ \ref{fig:KED_detectionAngle}b).

\begin{figure}[htbp]
  \centering
    \includegraphics[width=\textwidth]{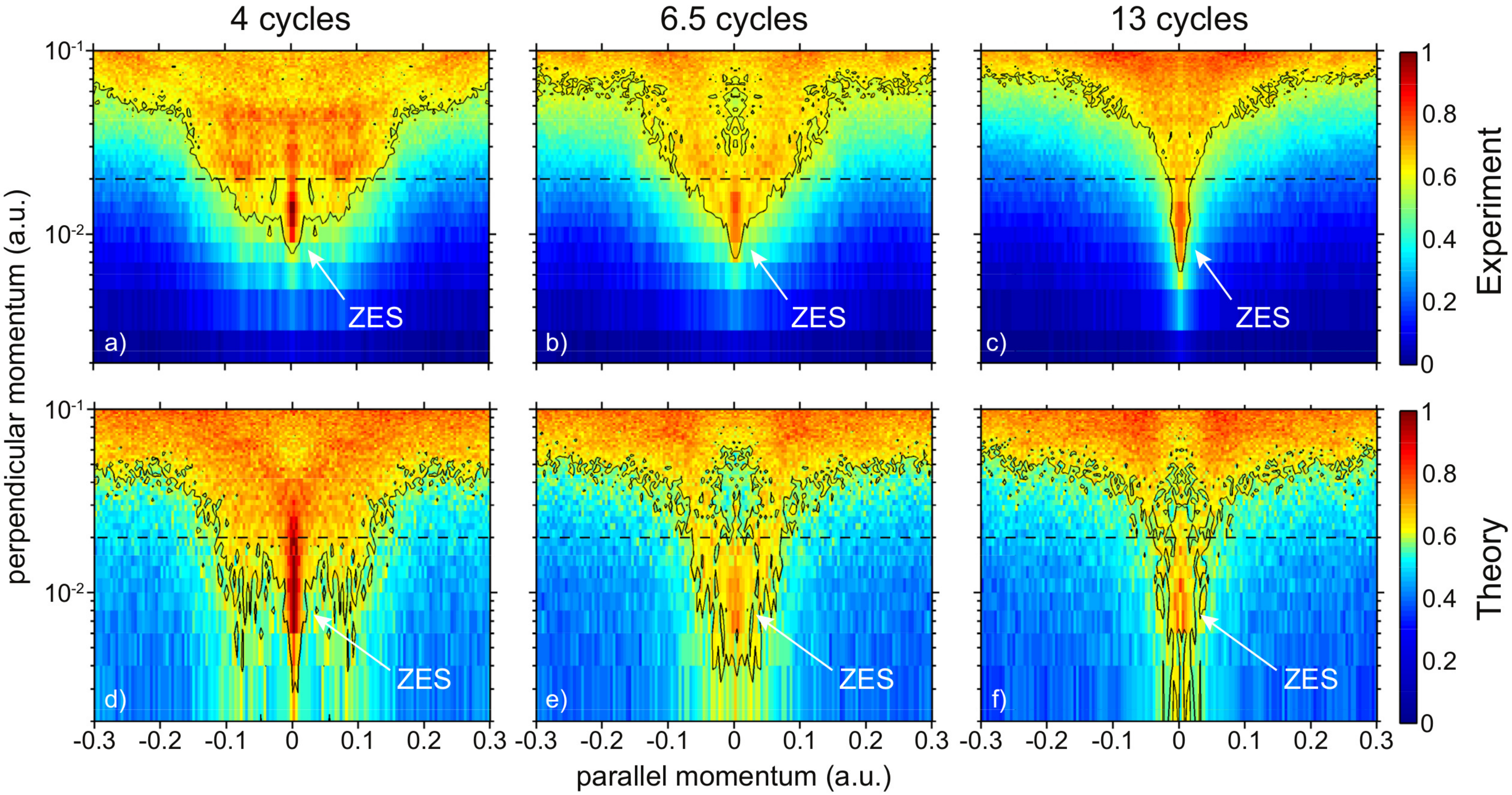}
  \caption{(Color online) Momentum map ($\log p_\perp$ over $p_\|$) of the measured (top) and simulated (bottom) emission spectra (normalized linear color scale) for different pulse durations $\tau_p=41$ fs (4 cycles, left), 68 fs (6.5 cycles, center), and 140 fs (13 cycles, right). The ZES is marked. A contour highlights the $60\,\%$ level  of the normalized data. The cut at $p_\perp=2\times 10^{-2}$ a.u.\ is used to determine the full width $\Gamma$ of the structures at very low energies.\label{fig:PtPl_log_zoom}}
\end{figure}

It is now instructive to study the variation of the LESs as a function of pulse duration, i.e., the number of laser field cycles. As focusing into the $k^{th}$ LES peak occurs for trajectories having turning points near the $z=0$ plane after the $k^{th}$ quiver oscillation following tunneling \cite{Lemell_LES}, increasing $k$ allows higher order LES and larger angular momenta. Indeed, as presented in Fig.\ \ref{fig:PtPl_log_zoom}, the VLES becomes increasingly sharper as $k$ or, equivalently, the pulse duration increases.
While the characteristic V-shape of the VLES is not yet recognizable for the shortest pulse duration $\tau_p=41$ fs, it develops for longer pulses. The reason for the absence of the V-shape is that for the shortest pulse duration (41 fs) only the first two orders of the LES can develop (compare Fig.\ 7 in \cite{Lemell_LES}).
\begin{figure}[h!]
	\centering
	  \includegraphics[width=0.40\textwidth]{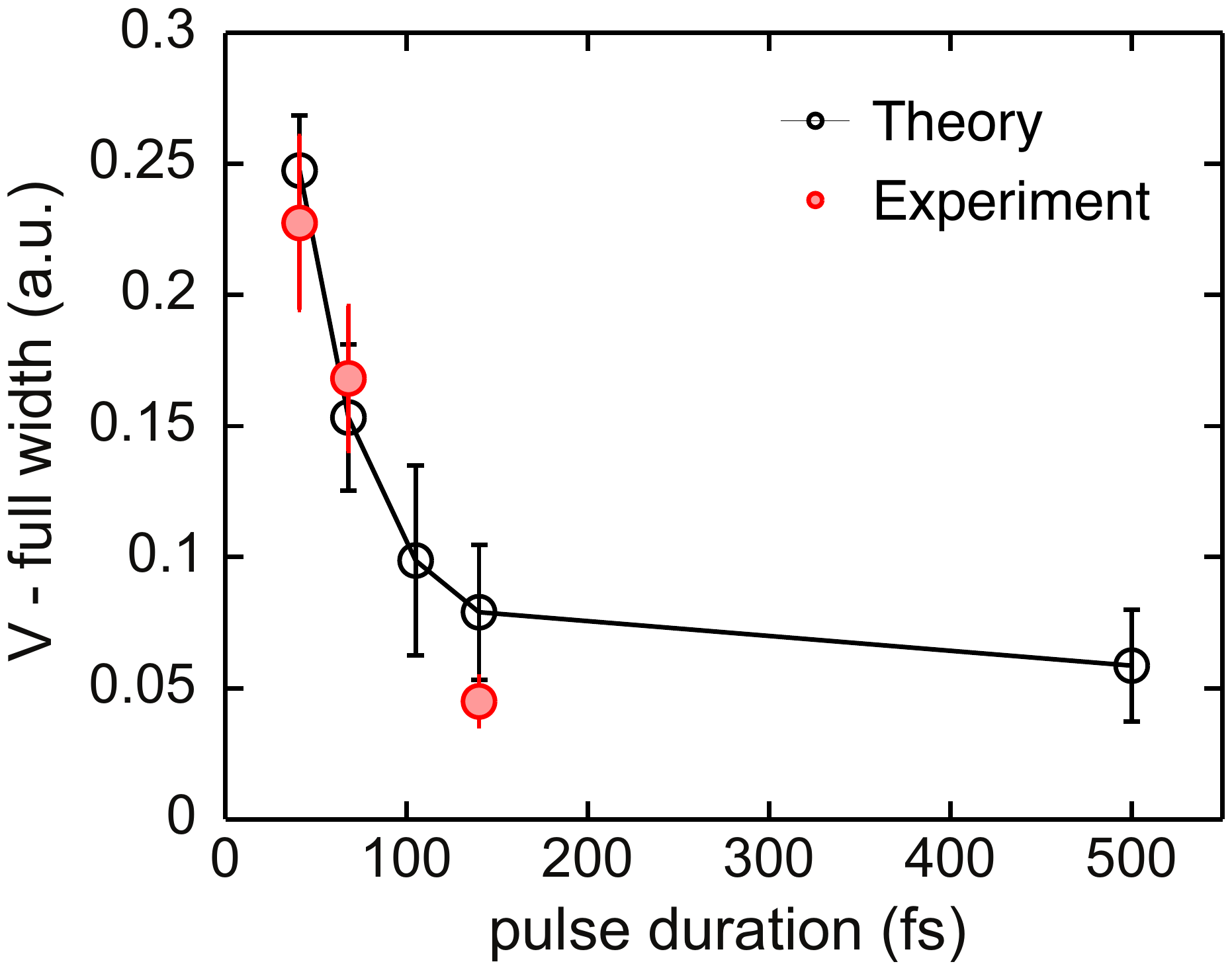}
	\caption{Full width $\Gamma$ of the structures at very low energies in $p_\|$ detected for a cut through the momentum distribution at $p_\perp=2\times 10^{-2}$ a.u.\
	intersecting with the $60\,\%$ level contour of the normalized data (see Fig.\ \ref{fig:PtPl_log_zoom}). Black line: CTMC simulation; red data points: experimental data.\label{JB1}}
\end{figure}
The total width of the structure reduces as $\tau_p$ gets longer. In Fig.\ \ref{JB1} the evolution of the full width $\Gamma$ of the structures at very low energies as a function of pulse duration is shown allowing for a quantitative comparison between experiment and simulation. The widths are extracted along the $p_\|$ direction taken at $p_\perp=2\times 10^{-2}$ a.u.\ (dashed lines in Fig.\ \ref{fig:PtPl_log_zoom}) intersecting with the $60\,\%$ level contour of the normalized data. The $60\,\%$ level was found to be the best compromise between signal statistics and the ability to resolve the structure. Both experiment and theory show a monotonic trend towards smaller width $\Gamma$ with increasing $\tau_p$. The agreement is quite remarkable given the uncertainty in the extraction of this observable and the small energies ($E<0.15$ eV) of the electrons in that momentum region which are treated within CTMC.

We now turn to the ZES, a feature found in all experimental momentum spectra of several atomic or molecular targets \cite{Dura_srep,Pullen}, with the local maximum appearing at $p_\|=0$ and $p_\perp\approx\Delta p_\perp$, where $\Delta p_\perp$ is the experimental momentum resolution perpendicular to the polarization axis. This peak is, within the experimental resolution, consistent with emission at zero kinetic energy and remains at the same position irrespective of the pulse duration. Such a structure is absent in the simulated emission spectrum when only direct ionization to the continuum is taken into account. However, the presence of a weak extraction field in the interaction volume of the ReMi, in the present case $F_\mathrm{extr}=1.3$ V/cm $\approx 2.6\times 10^{-10}$ a.u., allows for the extraction and therefore observation of electrons that end up in negative energy states $E<0$ relative to the ionization threshold of the atom at the end of the pulse. These transiently recaptured electrons can be field ionized by the extraction field of the ReMi long after the end of the pulse either by tunneling or over-barrier ionization. Approximating the ionic potential by a Coulomb tail and using hydrogenic Rydberg states, the threshold for over-the-barrier ionization above the saddle point at $E_\mathrm{bind}=-2\sqrt{F_\mathrm{extr}}$ lies near the principal quantum number $n=125$.

\begin{figure}[h!]
	\centering
	  \includegraphics[width=0.4\textwidth]{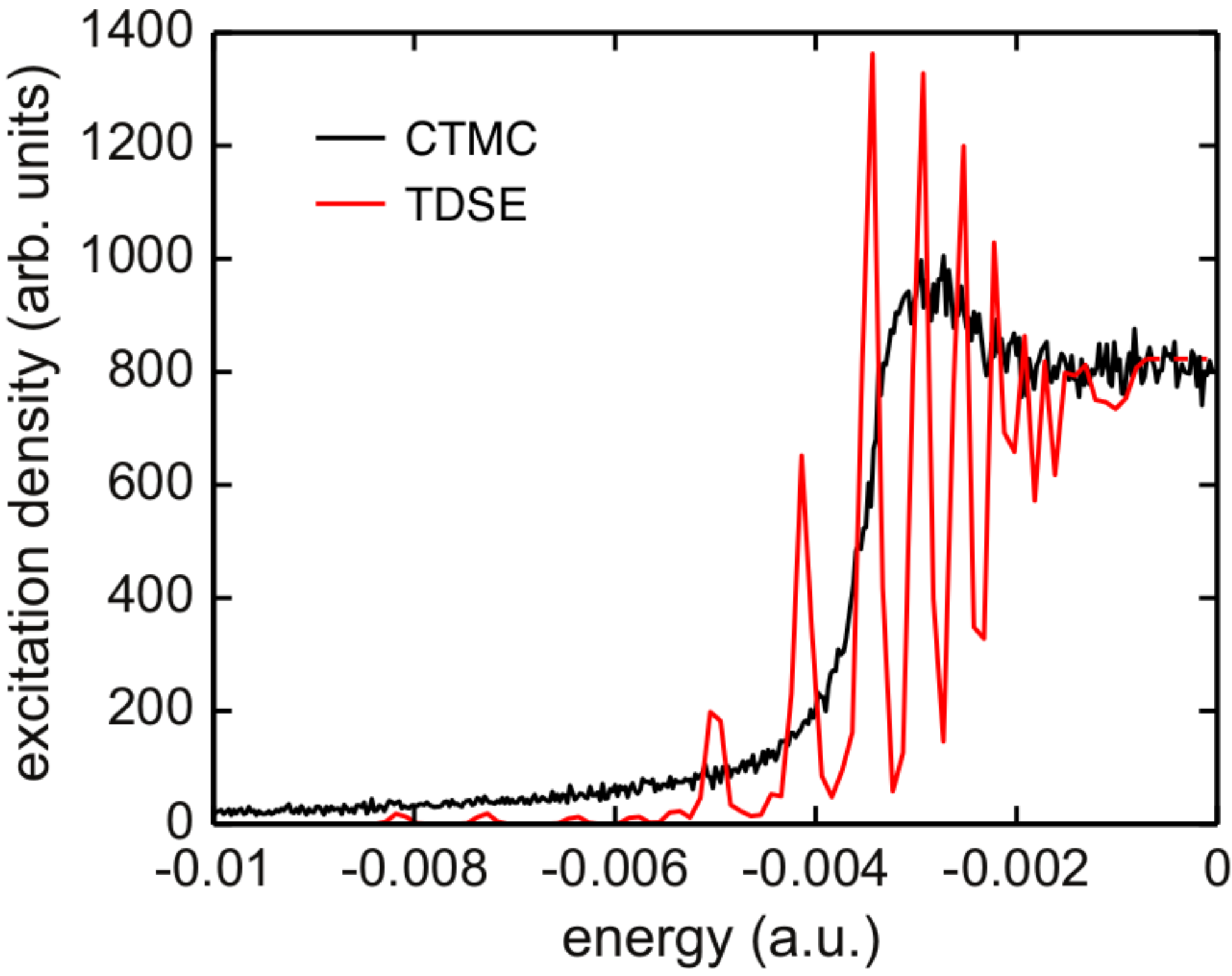}
	\caption{(Color online) Energy distribution of electrons after the end of a IR laser pulse (3100 nm, 68 fs, $10^{14}$ W/cm$^2$) from classical (black line) and quantum (red line) simulations.\label{occup}}
\end{figure}

There are two alternative pathways contributing to the population of such high Rydberg states: direct multi-photon like excitation from the ground state and indirectly through recapture of tunnel-ionized electrons. The latter process has been referred to as frustrated field ionization \cite{Nubbe,Eich} and has been unambiguously identified for Rydberg atoms exposed to a sequence of ultrashort pulses \cite{Yosh}. Since only the second pathway is accounted for by the present CTMC simulation, probing the spectral distribution below the ionization threshold allows for the determination of the relative importance of the two excitation processes. Moreover, since, unlike the CTMC, the quantum simulation contains all pathways, this is also a sensitive probe of the classical--quantum correspondence. In Fig.\ \ref{occup} the energy distribution of electrons after the end of a IR laser pulse (3100 nm, 68 fs, $10^{14}$ W/cm$^2$) is plotted for the quantum and classical simulations. Both results feature a steep rise near $E=-0.0035$ a.u.\ and a nearly constant spectral excitation density up to the ionization threshold.
The onset of Rydberg population is near $n\gtrsim 12$. The existence of this threshold in both the quantum and classical simulations is a clear indication that recapture of tunnel-ionized electrons dominates over direct bound state excitation. For the latter such a sharp onset is not expected. Moreover, the value for the threshold can be easily estimated from the recapture scenario. Comparing the Coulomb potential of an electrons with vanishing kinetic quiver energy at a distance corresponding to the quiver radius $\alpha$ at the end of the pulse with the (hydrogenic) binding energy of a Rydberg state with principal quantum number $n_c$ gives
\begin{equation}
-\frac{1}{\alpha}=-\frac{1}{2n_c^2}
\end{equation}
yielding $n_c\approx 11$ and $E=-0.004$ remarkably close to the onset in Fig.\ \ref{occup}. The CTMC simulation is thus well suited to describe the recapture process and the ZES emerging from it. This is particularly convenient since the spectral region close to threshold contributing to field ionization ($n\gtrsim 125$) cannot be easily accurately represented by the discrete pseudostate expansion in the quantum simulation.

Within the CTMC the relative weight of the ZES compared to the surrounding low-energy spectrum can be easily determined. Since both contributions originate from the same source, we compare them to the total number $N_I$ of the ensemble of tunnel ionized electrons during the pulse. From those, the fraction
\begin{equation}
P_\mathrm{low-E}=N_\mathrm{low-E}/N_I
\end{equation}
contributes to the smooth low-energy spectrum defined here by electrons with positive energy and final momenta $p\leq 0.1$ a.u. The fraction of recaptured electrons
\begin{equation}
P_\mathrm{ZES}=N_\mathrm{rec}/N_I\, ,\label{eq9}
\end{equation}
where $N_\mathrm{rec}$ is the number of electrons captured into weakly bound Rydberg states above the field-ionization threshold, contribute to the ZES. For a pulse duration of 68 fs the ratio $F_\mathrm{ZES}=P_\mathrm{ZES}/(P_\mathrm{ZES}+P_\mathrm{low-E})$ for the CTMC results gives $F_\mathrm{ZES}=0.006$ very close to the estimate of $F_\mathrm{ZES}=0.0063$ assuming constant excitation density $\rho(E)$ across threshold. A weakly increasing function $\rho(E)$ as found in the simulation gives rise to a slightly smaller simulated value of $F_\mathrm{ZES}$.

In the experiment, ZES electrons cannot be distinguished from the low-energy background, requiring a modification of Eq.\ \ref{eq9}. We, instead, determine the number of electrons $N_\mathrm{ZES}$ with asymptotic momentum $p\leq 0.018$ a.u.\ and evaluate $F_\mathrm{ZES}=N_\mathrm{ZES}/N_\mathrm{low-E}$. For the comparison with the simulation we fold the simulated distribution of recaptured electrons with a Gaussian with widths $\sigma_\| =\Delta p_\|$ and $\sigma_\perp =\Delta p_\perp$ given by the experimental resolution centered near the origin $(p_\|=0, p_\perp=\Delta p_\perp)$ and add them to the momentum distributions (marked as ZES in Fig.\ \ref{fig:PtPl_log_zoom}d, e, f). We consider only over-barrier ionization neglecting tunnel ionization of Rydberg states with principal number smaller but close to $n\lesssim 125$. Assuming again constant $\rho(E)$ around threshold we arrive at the estimate $F_\mathrm{ZES}\geq 0.0324$ which fits well with the simulated values of $F_\mathrm{ZES}=0.038$ for 41 fs pulses slowly decreasing to $F_\mathrm{ZES}=0.036$ for 140 fs pulses. The experimental value for $F_\mathrm{ZES}$ is slightly smaller ($F_\mathrm{ZES}=0.025$ for 41 fs pulses, $F_\mathrm{ZES}=0.027$ for 140 fs pulses) which might be directly related to the reduced detection probability for electrons with asymptotic momenta below the perpendicular momentum resolution of the ReMi, $p<\Delta p_\perp$. Overall, the \textit{quantitative} agreement between simulation and experiment is remarkable.

\section{Summary}
We have presented a joint experimental and theoretical study analyzing higher order low-energy structures in strong-field ionization by ultrashort mid-IR laser pulses. We have shown that the V-shaped structure in the vectorial momentum distributions at very low energies ($E<0.5$ eV), referred to as ``very-low energy structure'' (VLES) is due to high-order focal points in the combined Coulomb and laser fields. Changes of structures in the vectorial momentum distributions reflect the two-dimensional focusing with increasing order most clearly visible when investigated in the energy--angular momentum plane. We show that the VLES shifts towards larger angular momenta with increasing pulse length leading to a narrowing of the V-shaped structure in momentum space. We have also identified the zero-energy structure (ZES) as high-lying Rydberg states field ionized by the extraction field of the ReMi. The ZES provides, thus, quantitative information on recapture of tunnel-ionized electrons and represents a map of the Rydberg population onto the momentum plane. We have shown that the quantum-classical correspondence holds also for this low energy portion of the electron spectrum.

\acknowledgments{This work was supported by MINISTERIO DE ECONOMA Y COMPETITIVIDAD through Plan Nacional (FIS2011-30465-C02-01), the Catalan Agencia de Gestio d'Ajuts Universitaris i de Recerca (AGAUR) with SGR 2014-2016, Fundacio Cellex Barcelona, funding from LASERLAB-EUROPE, grant agreement 228334, and by the FWF (Austria) as part of the special research project SFB-049 Next Lite. B. W. was supported by AGAUR with a pre-doctoral grant (FI-DGR 2013-2015).
X.-M. T. was supported by a Grant-in-Aid for Scientific Research (Grant No.\ C24540421) from the JSPS. M. G. P. was supported by a Marie Curie ICFOnest co-funding fellowship. J. B. and J. B. acknowledge partial support by NSF grant PHY II-25915.}

\end{document}